\documentclass{aa}
\usepackage{natbib}
\usepackage{graphicx}
\usepackage[mathcal]{eucal}
\usepackage{amssymb} 

\newcommand{\ds}{\displaystyle}

\long\def\jumpover#1{{}}

\newcommand{\derivp} [2] {\frac {\partial #1 } {\partial #2} }

\newcommand{\fig}[3]{
      \begin{figure}[ht]
        \begin{center}
        \resizebox{\hsize}{!}{\includegraphics  {#1}}
        \end{center}    
        \caption{#2}
        \label{#3}
        \end{figure} }
\newcommand{\eqn} [1] {
\begin{equation}#1
\end{equation}}

\newlength{\lenA} %
\begin{document}

\title{Influence of local  treatments of convection 
 upon solar $p$~mode excitation rates.}
\author{Samadi R. \inst{1,2}    \and Kupka F. \inst{3} \and Goupil M.J. \inst{1}   \and Lebreton Y. \inst{4}  \and van't Veer-Menneret C. \inst{4} } 
\institute{
Observatoire de Paris, LESIA, CNRS UMR 8109, 92195 Meudon, France \and 
Observat\'orio Astron\'omico UC, Coimbra, Portugal \and
Max-Planck-Institute for Astrophysics, Karl-Schwarzschild Str. 1, 85741 Garching\and 
Observatoire de Paris, GEPI, CNRS UMR 8111, 92195 Meudon, France
}
\offprints{R. Samadi}
\mail{Reza.Samadi@obspm.fr}
\date{\today} 

\titlerunning{.}

\abstract{ 
We compute the rates $P$ at which acoustic energy is injected into the solar 
radial $p$~modes for several solar models. The solar models are computed with two
different local treatments of convection: the classical mixing-length theory (MLT hereafter)
and \citet[][ CGM96 hereafter]{Canuto96}'s formulation. Among the models investigated here,
our best models reproduce both the solar radius and the solar luminosity at solar age
\emph{and} the observed Balmer line profiles. For the MLT treatment, the rates $P$ do
depend significantly on the properties of the atmosphere whereas for the CGM96's treatment
the dependence of $P$ on the properties of the atmosphere is found smaller than the error
bars attached to the seismic measurements. The excitation rates $P$ for modes associated with the MLT
models are significantly  underestimated compared with the solar seismic constraints.
The CGM models yield values for $P$ closer to the seismic data than the MLT models. We
conclude that the solar p-mode excitation rates provide valuable constraints and according
to the present investigation clearly favor the CGM treatment with respect to the MLT,
although neither of them yields values of $P$ as close to the observations
as recently found for 3D numerical simulations.
}
\maketitle

\section{Introduction}

In the  outermost part of the convective zone (CZ) of intermediate mass stars,  
convection is highly superadiabatic because of the rapid radiative heat gains and losses 
of the convective fluid. In that region, entropy fluctuations are the largest and 
the resulting decrease of the convective transport efficiency  is compensated
 by a large increase of the eddy motions which is responsible for the oscillation mode driving.
Modelling inefficient convection is complex. 3D numerical simulations are now being  performed 
but  remain  still very time consuming. Hence  for massive stellar computations, 
1D stellar models are used in which only simple prescriptions of convection
are implemented.

Among  these simplified treatments,  \citet[][ CM91 hereafter]{Canuto91}'s approach
differs  from the classical mixing length approach (MLT hereafter) in that it takes 
into account the contribution of eddies with different sizes  in the calculation of 
the convective flux and velocity while keeping the computational expenses as low as the MLT.
An improved version was proposed by \citet[ CGM96 hereafter]{Canuto96} which   takes into account 
the feedback of the turbulence on the energy input from the source which generates turbulent convection.
These multi-eddy convection models are usually refered to as Full Spectrum of Turbulence
(FST) models.

Several non-local formulations of convection have also been proposed
\citep{Gough77,Xiong78,Xiong85,Canuto92,Canuto93,Canuto98}. However, we focus
here on the effects of proposed improvements in the description of the energy
spectrum and therefore consider only local treatments and compare
FST models with MLT ones.

Any model of convection must satisfy several observational constraints provided
by our Sun: the solar radius at the solar age, the Balmer line profiles 
and the $uvby$ color indices. The MLT, CM91's and CGM96's local treatments
have been confronted to these observational constraints 
\citep[e.g.][]{Fuhrmann93,Fuhrmann94,Vantveer96,Smalley97,Bern98,Heiter02,Mont04}.
One main result is that these observational quantities are more sensitive to
the adopted value of the convective scale length of the eddies than to the
formulation of convection.

Solar seismic observations provide strong additional constraints. Comparisons 
of theoretical oscillation frequencies with observed solar ones have shown for
instance that significant improvement in the agreement between observation and
model at high frequency and degree $\ell$ can be achieved with 3D simulations
\citep{Rosenthal99}. We are here interested in amplitudes of oscillation which
can also bring several constraints on the convective process in the outer
solar envelope. Indeed, the amplitudes of solar-like oscillations result from
a balance between excitation and damping.
Measurements of the oscillation mode growth rates (through their line-widths) 
and of the mode amplitudes enable the evaluation of the excitation rates $P$.
Excitation of solar-like oscillations is known to be both due to turbulent convective
motions through the driving by the turbulent Reynolds stresses
\citep[see][]{GK77,Balmforth92c,Samadi00I} and due to the advection of turbulent
entropy fluctuations by the turbulent movements \citep[see][]{Samadi00I}.
The excitation rates $P$ are thus directly related to the velocity of the 
convective elements and to the amount of thermal energy advected by convective motions  
(i.e.\ the convective flux). The excitation rates then crucially depend on the way 
the convective velocity and flux are modelled \citep[see][]{Houdek99}.
Solar seismic measurements therefore provide -- through a model of mode excitation -- 
additional constraints on the stellar convective properties. In this framework, the goal
of the present paper is to investigate the influence of different local treatments of
convection on the calculation of the rates at which energy is injected into the solar
radial $p$~modes and to compare our results with the solar seismic constraints. 

For this purpose we compute two calibrated solar models with the B\"ohm-Vitense 
formulation of the MLT \citep[][ hereafter BV]{Bohm58} and with the  CGM96
multi-eddy convection treatment. In each case, the same convection formulation is
adopted for the interior {\it and} the model atmosphere. Models for the internal structure
are built so as to reproduce the solar radius and the solar luminosity at the solar
age. The atmosphere of each model is constructed using a $T(\tau)$ law which is derived
from a Kurucz's model atmosphere \citep{Kurucz93} computed with the same convection
formulation \citep[as described in][]{Heiter02}. These model atmospheres are built in
order to provide the best agreement between synthetic and observed Balmer line profiles
  \citep[as in][ for the MLT treatment]{Vantveer96} (Sect.~2).
The matching of the model atmosphere with the interior model is performed 
-- in the manner of \citet{Morel94} -- by ensuring the continuity of the temperature
gradient, $\nabla$, and of the convective flux in a transition region between the
interior and the atmosphere.

We compute also two models with an Eddington gray atmosphere. 
One with the MLT treatment and the second one with the CGM96 formulation.
These two additional models are considered for comparison purpose only.
Indeed, they have an atmosphere with the same mixing-length parameter as in 
the interior and do not reproduce the Balmer line profiles.
As a consequence, in contrast with the interior models including a Kurucz's 
atmosphere as described above, their  atmospheres do not fulfill constraints 
on the properties of the convection at the surface.

Calculation of the excitation rates requires the computation of the 
convective flux, $F_{\rm c}$, and of the convective velocity, $v$. This 
is done in Sect.~3 by paying special attention to the problem of the 
transition region. Indeed, the continuity of $\nabla$ and of $F_{\rm c}$ 
through the transition region imposes a spatial variation of the mixing-length
parameter in the transition region. This variable mixing-length parameter is then
used in Sect.~3 to compute $v$. Note that our approach, which is used here to
compute $v$ and $P$, is different from that of \citet{Schlattl97} who built
stellar models assuming a spatially varying mixing-length parameter, with
a spatial variation imposed a priori from a comparison to 2D numerical simulations
of convection, in order to compute $p$~mode frequencies.

As a last step (Sect.~4), we compute the adiabatic eigenmodes  and the excitation rates $P$   for each model. 
The adopted model of excitation is that of \citet[ Paper~I hereafter]{Samadi00I} in which 
the characteristic wavenumber $k_0$, the wavenumber dependency of the turbulent spectra 
as well as the frequency component ($\chi_k$) of the  correlation product  
of the turbulent velocity field are constrained with a 3D simulation of the Sun 
as in  \citet[ Paper~II hereafter]{Samadi02I} and  \citet[ Paper~III hereafter]{Samadi02II}. 
Comparison with  solar seismic constraints then allows 
to conclude about the best local treatment of convection 
in the solar case  (Sect.~5).

\section{Solar models}
\label{sec:Matching interior models to  model atmospheres}

All solar models discussed here are computed with the CESAM code \citep{Morel97} including
the following input physics and numerical features:
\begin{enumerate}
\item Equation of state (EOS): CEFF EOS \citep{JCD92}.
\item Opacities: OPAL \citep{Iglesias96} data
complemented by \citet{Alexander94} data for
$T\lesssim10^4\ \rm K$, both set of data being given for \citet{GN93} solar mixture.
\item Thermonuclear reaction rates: \citet{Caughlan88}.
\item Convection: either MLT or CGM96's formalism.
The same convection formalism has been used in the interior and in the model atmosphere.
\item Microscopic diffusion: all models include microscopic diffusion of helium and
heavy elements calculated according to the simplified formalism of  \citet{Michaud93}
where heavy elements are treated as
trace elements.
\item Chemical composition and mixing length parameter for convection: 
the \citet{GN93} heavy elements solar mixture has been adopted.
The constraint that solar models have  the observed solar luminosity and radius at solar age
yields the initial helium content $Y_0$ and the mixing length parameter of the interior model
$\alpha_{\rm i}$ (solar model calibration). Microscopic diffusion modifies the 
surface composition, therefore
the initial ratio of heavy elements to hydrogen $(Z/X)_{\rm 0}$ is adjusted so as to get the  
ratio $(Z/X)_{\odot}=0.0245$ at solar age. 
\item The models were calculated with 285 shells in the atmosphere and about 2000 shells
in the interior.
\end{enumerate}

The CGM96 formulation of convection is implemented according to \citet{Heiter02}.
In contrast with \citet{Heiter02}, we use for the two formulations a characteristic scale
length of convection which is the mixing-length $\Lambda = \alpha \, H_p$ where $H_p$ is
the pressure scale heigth and $\alpha$ is the mixing-length parameter which can be different
in the interior and in the atmosphere.

\subsection{The ``Kurucz models'' (KMLT and KCGM models)}
\label{sec:KMLT and KCGM models}

We consider here two stellar models: one computed with the MLT formulation of
convection and the second one with the CGM96 formulation.
They will be hereafter labelled as KMLT model and KCGM model respectively.

\begin{figure*}
        \resizebox{\hsize}{!}{\includegraphics[angle=-90]{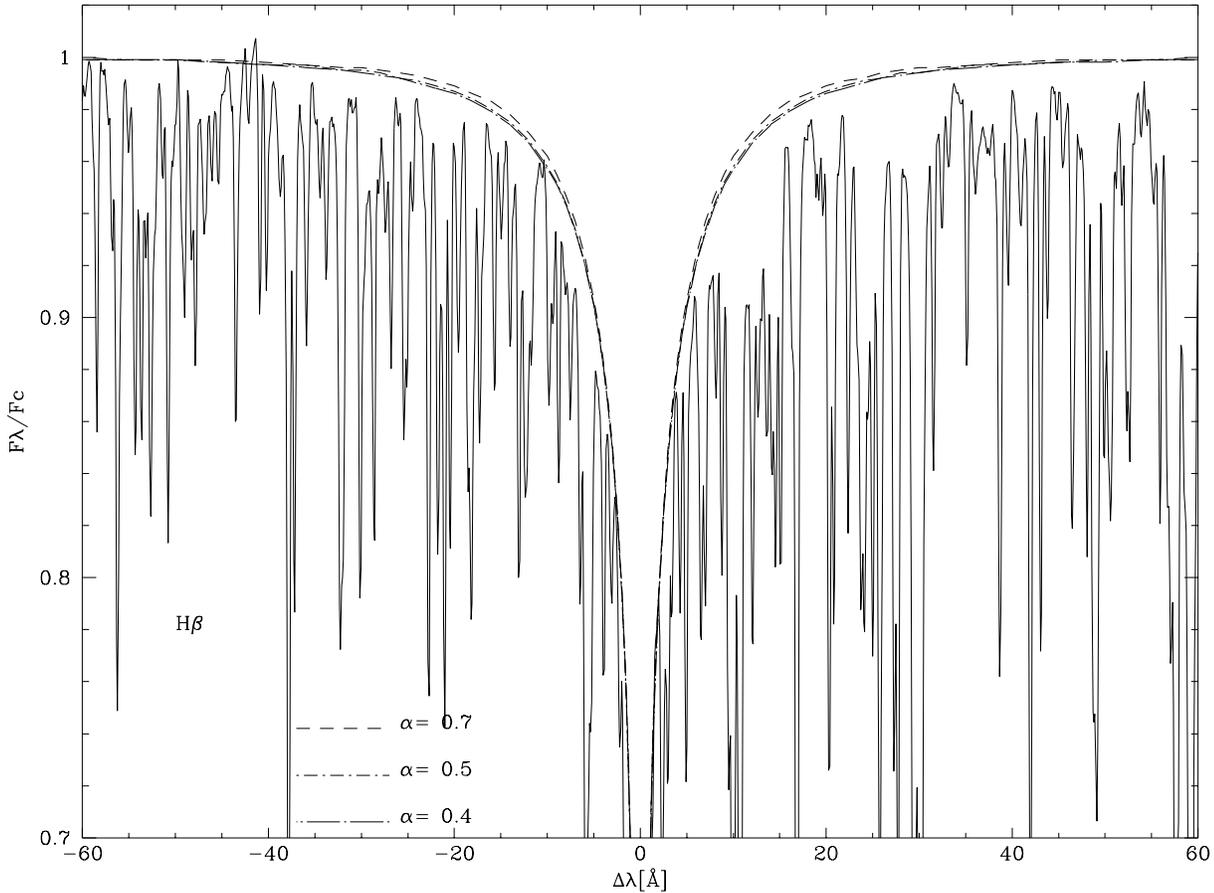}}
        \caption{ The observed solar ${\rm H}_\beta$ profile is compared to theoretical ones 
computed with CGM models and assuming different values for $\alpha_{\rm a}$.  
      Abscissae are distances in {\AA} from the line center,
 ordinates are the flux in the profile normalized to the continuum. The large
 scatter is due to the presence of many  spectral lines which
 overlap and cause an apparent enlargement of the true profile. 
}
        \label{fig:HB_CGMB}
        \end{figure*}

{\it Treatment of the atmosphere:}
The model atmospheres of those models are computed using the ATLAS~9 code
\citep{Kurucz93} as described in \citet{Heiter02}. 
Solar model atmospheres  are built assuming different values for  $\alpha_{\rm a}$, the
mixing-length parameter assumed for the model atmosphere: $\alpha_{\rm a}=0.4, 0.5,
0.6, 0.7 $. The model atmospheres with $\alpha_{\rm a}=0.4$ and $\alpha_{\rm a} = 0.5$
provide the  best agreement between synthetic and observed ${\rm H}_\beta$ Balmer line
profiles for the two formulations of convection. 
This is shown in  Fig.~\ref{fig:HB_CGMB} for the CGM model atmosphere.
For the MLT treatment see \citet{Fuhrmann93,Fuhrmann94} and \citet{Vantveer96}.  
Above $\alpha_{\rm a} \simeq 0.6$, the synthetic profile rapidly departs from the observed
one as well as the effective temperature $T_{\rm eff}$ from the known solar $T_{\rm eff}$. 

There are no significant differences for the ${\rm H}_\beta$ Balmer line profile
between the  model atmospheres with $\alpha_{\rm a}=0.4$ and $\alpha_{\rm a} = 0.5$.
Among those model atmospheres we adopt arbitrary those with $\alpha_{\rm a} = 0.5$.
Indeed, choosing the model atmospheres with $\alpha_{\rm a}=0.4$ instead of
$\alpha_{\rm a}=0.5$ will not change the conclusions of this article.

For each formulation of convection we then obtain a $T(\tau)$ law.

The atmospheres of the KMLT and KCGM stellar models are  recomputed  according to the
procedure described in \citet{Morel94} from the $T(\tau)$-laws  mentioned above: the
fit between interior (where the diffusion approximation is valid) and atmosphere is
performed in a region where $\tau_1 \la \tau \la \tau_2$ (an acceptable range of values
for $\tau_1$ and $\tau_2$ is discussed below). In the interior region, where $\tau \ga 
\tau_2$, the temperature gradient $\nabla_{\rm i}$ is obtained from the MLT or CGM96
formalism. In the atmospheric region, where $\tau \la \tau_1$, the temperature gradient
$\nabla_{\rm a}$ is computed using the $T-\tau$ law of the model atmosphere built with
the same model of convection as in the interior. In the transition region, where
$\tau_1 \la \tau \la \tau_2$, in order to ensure the continuity of the temperature
gradient, $\nabla$ is obtained by a linear interpolation of $\nabla_{\rm i}$ and
$\nabla_{\rm a}$ as a function of the optical depth as follows:
\eqn{
\nabla = \beta(\tau) \nabla_{\rm a} + (1-\beta(\tau)) \nabla_{\rm i}
\label{eqn:nabla:full_model}
}
where $\beta(\tau) =\ds  (\tau_2-\tau)/(\tau_2-\tau_1)$.

Once the temperature gradients of the interior and the atmosphere are linked together in the transition region according to Eq.~(\ref{eqn:nabla:full_model}), we compute \emph{afterward} in that region the convective flux and an \emph{equivalent} mixing-length parameter (i.e. a depth dependent mixing-length parameter)  as explained in Sect.~3.

\medskip

{\it Acceptable ranges for $\tau_1$ and $\tau_2$:}
Using a Newton-Raphson scheme, $\nabla_{\rm i}$ is adjusted in order that
$F_c^{(\rm i)}+F_{\rm rad}^{(\rm i)} = L_\odot/ 4\pi r^2$, where $L_\odot$ is the
luminosity of the Sun and $F_{\rm rad}^{(\rm i)}$ is the radiative flux of the internal
model. Calculation of $F_c^{(\rm i)}$  assumes for  $F_{\rm rad}$ the diffusion
approximation for the radiative transfer. This approximation is valid at rather high
values of $\tau$, typically $\tau \gtrsim 10$ \citep[see][]{Morel94}. Therefore,
$\tau_1$ cannot be much smaller than $\tau \simeq 10$. Otherwise, $\nabla_{\rm i}$
will have an unrealistic contribution to $\nabla$ below $\tau \simeq 10$. On the other
hand, for the calculation of $F_c^{\rm a}$, the radiative flux  $F_{\rm rad}$ is based
on a Kurucz's model atmosphere which treats the radiative transfer more realistically
than the diffusion approximation.

The Kurucz model atmosphere is based on \citet{Kurucz92,Kurucz93} opacity tables
which are given up to $T \simeq 2\cdot 10^5$~K and $P = 10^8$ dyn~cm$^{-2}$.
As a consequence, $\tau_2$ cannot be larger than $\tau \simeq 10^{7.5}$, i.e.\
layers for which $T \simeq 30000$~K. In order to ensure a satisfactory
continuity of the temperature gradient $\tau_2$ must be sufficiently larger
than $\tau_1$. On the other hand, the transition region should be as small as
possible, i.e.\ $\tau_2-\tau_1$ must be as small as possible. The main
constraint for this region is thus to avoid discontinuities between the
interior and the atmosphere. It is defined through an empirical
procedure rather than based on a strict physical theory.

In practice, we find that $\tau_1=4$ is the minimal acceptable value for $\tau_1$;
below this value the bias introduced by the diffusion approximation has a significant
effect on $F_{c}$. In addition, we find that above $\tau_2 \simeq 50$, the convective
velocity $v$ (see Sect.~\ref{sec:Convective velocity}) shows a pronounced
 ``kink'' at $\tau=\tau_2$ for the CGM model (see Fig.~\ref{fig:fc:alpha:v}, such
a  ``kink'' is also observed for $F_c$, but it is less pronounced). On the other
hand, the choice of $\tau_2=20$ avoids the angular point. For the MLT model,
whatever the value of $\tau_2$, $v$ shows such a feature. This is a consequence
of the much larger values of the mixing length parameter required in the interior
in order to still obtain the correct solar radius, if a lower value of $\alpha$
is used also in regions further within the envelope. The requirement of matching
$R_\odot$ hence provides a more stringent upper limit for the choice of $\tau_2$
(cf.\ also the discussion in \citealt{Mont04} on the computation of solar entropy
as a function of radius). In the following, we
will consider $\tau_1=4$ and $\tau_2=20$ as our optimal choice.

{\it Calibration:}
The mixing-length parameter $\alpha_i$ for the internal structure, $Y_0$, and $(Z/X)_0$
are adjusted such that the stellar model simultaneously reproduces the solar radius,
the  solar luminosity, and the observed ratio $(Z/X)_{\odot}=0.0245$ at the solar age. 
The calibration yields $(Z/X)_{\rm 0}=0.0279$ and $Y_0=0.275$. At solar age, the helium
abundance in the convection zone is $Y=0.246$ in reasonable agreement with the value
$Y=0.249\pm0.003$ obtained from seismology \citep{Basu97}. Table~\ref{tab:kurucz} 
gives the calibrated values of the mixing-length parameters $\alpha_{\rm i}$ for each complete solar  model.  
The radius resulting from the adjustement of $\alpha_{\rm i}$ as well as the size of
the convective zone are given in Table~\ref{tab:radius:depth CZ}. All the interior
models have a depth of the convective zone of $\simeq 0.286~R_\odot$ which is in 
good agreement with the value of $0.287\pm0.003\,R_\odot$ 
determined seismically by \citet{JCD91}. 

\begin{table}[b]
\caption{Values of the mixing-length parameters of the KCGM and KMLT models:
$\alpha_{\rm i}$ (for the interior) and $\alpha_{\rm a}$ (for the model atmosphere).
$\alpha_{\rm i}$ results from the calibration of the full model while $\alpha_{\rm a}$
is fixed (see Sect.~2.1).
}
\label{tab:kurucz}
\begin{center}
\begin{tabular}{cccc}  
model  &$\alpha_{\rm i}$ & $\alpha_{\rm a}$\cr
\hline
KMLT & 2.51  & 0.50 \cr
KCGM & 0.78 & 0.50 \cr 
\hline
\end{tabular}
\end{center}
\end{table}

\begin{table}[b]
\caption{$\Delta R \equiv R- R_\odot$, where $R_\odot$ is the radius at the
photosphere ({\it i.e.} at $T=T_{\rm eff}$), and depth of the convective zone
(CZ) for the KCGM and KMLT models. These quantities are given with respect to
the solar radius $R_\odot$ (we assume \citet{Guenther92}'s value of $R_\odot$).}
\label{tab:radius:depth CZ}
\begin{center}
\begin{tabular}{lll}  
model &$\Delta R/R_\odot$ & depth CZ\cr
\hline
KMLT &  $-10^{-6}$ & 0.2860\cr
KCGM &  $5~10^{-6}$ & 0.2859 \cr
\hline
\end{tabular}
\end{center}
\end{table}

\subsection{Eddington approximation based models (EMLT and ECGM models)}

For comparison purpose, we consider here two additional stellar models with an
Eddington classical gray atmosphere instead of the Kurucz atmosphere models described
in Sect~\ref{sec:KMLT and KCGM models}. One of these models assumes the MLT formulation
of convection and the other one the CGM96 formulation. In the following they will be
labelled as EMLT model and ECGM model, respectively. The mixing-length parameter
$\alpha$ of these models (the same $\alpha$  in the interior as in the atmosphere)
is adjusted in order to reproduce the solar luminosity and radius at the solar age.
However, as mentioned in the introduction, these models do not reproduce the Balmer
line profiles. Table~\ref{tab:edd} gives the calibrated values of the mixing-length
parameters.

\begin{table}[b]
\caption{Values of the the mixing-length parameter $\alpha$ of the ECGM and EMLT models
obtained for calibrated solar models.
}
\label{tab:edd}
\begin{center}
\begin{tabular}{cccc}  
model &$\alpha$ & \cr
\hline
EMLT & 1.76     \cr
ECGM & 0.69      \cr
\hline 
\end{tabular}
\end{center}
\end{table} 

\subsection {Comments}
\label{sect:The solar models:Comments}

With the CGM96 treatment,  $\alpha_{\rm i}$ is found less than one 
and closer to $\alpha_{\rm a}=0.5$. 
In contrast, with the MLT treatment the value of  $\alpha_{\rm i}$ 
is much larger than $\alpha_{\rm a}=0.5$. 

The CGM models result in a much smaller value for the mixing-length parameter
($\alpha_{\rm i}$ for the KCGM model and $\alpha$ for the ECGM model) than the
MLT models because the convection in nearly adiabatic regions is more  efficient  
with the CGM96 formulation than with the MLT one.  Indeed, for the same value of the
mixing-length parameter  and in the region of high convective efficiency (below the
superadiabatic region), the CGM96 treatment predicts a convective flux ten times
larger than the MLT one for a given superadiabatic gradient. In a solar model
this behavior results in a gradient closer to the adiabatic one below a smaller
superadiabatic zone in comparison to the MLT case \citep{Canuto96}.
At the top of the quasi-adiabatic region, energy is predominantly
transported by convection, such that $F_c \simeq F $ where  $F_c$ is the convective
flux and $F$ is the total energy flux. Above, convective transport is no longer
efficient ($F_c < F$). Therefore, the solar energy flux $F$ at the top of the
quasi-adiabatic region can be reproduced for the CGM models with a smaller value
of the mixing-length parameter than that one required for the MLT.

\begin{figure}[ht]
\begin{center}
\includegraphics[width=\lenA] {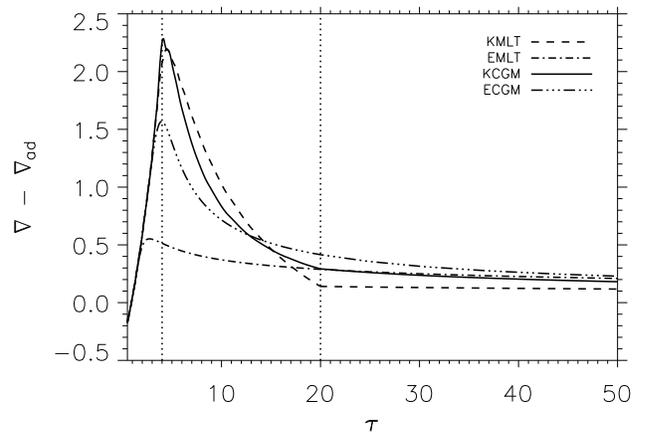}
\end{center}    
\caption{The superadiabatic gradient $\nabla-\nabla_{\rm ad}$ is plotted versus the
  optical depth $\tau$ in the outer region for the KMLT model (dashed line), EMLT model
  (dot-dashed line), ECGM model (dot-dot-dot-dashed line) and KCGM model (solid line).
  The dotted vertical lines correspond to the layers where $\tau=\tau_1=4$ and
  $\tau=\tau_2=20$, respectively, and delimit the transition region.
}
\label{fig:grad:v:fc}
        \end{figure}

The superadiabatic gradients are displayed in Fig.~\ref{fig:grad:v:fc} as a
function of $\tau$. The angular point observed in $\nabla-\nabla_{\rm ad}$ at the
optical depth $\tau \simeq 20$ for the KMLT model corresponds to the matching
point ({\it i.e.} at $R_\odot - r \simeq 100$~km, where $R_\odot$ is the solar
radius at the photosphere, defined to be where $\tau=2/3$ and calibrated at the
precision level given in Table~\ref{tab:radius:depth CZ}). It results from the
large difference between $\alpha_{\rm i}$ and  $\alpha_{\rm a}$. This difference
is much smaller for the KCGM model and therefore the  ``kink'' at
$\tau \simeq 20$ is much less pronounced.

 Our last comment concerns the large difference in the value of the mixing-length parameter between the KMLT and the EMLT models:
Both models differs only by the structure of their upper most layers located 50 - 100km below the photosphere  (see Fig.~3, middle) which represent a tiny fraction of the convection zone depth. The KMLT has a model atmosphere in which convection is much less efficient than that of the EMLT model as a consequence of the fact that ${\alpha_{\rm a}=0.5}$ in the Kurucz's model atmosphere (see Sec.~2.1). This is the reason why the superadiabatic gradient, $\nabla-\nabla_{\rm ad}$, is reaching  at that depth much higher values for the KMLT model than for the EMLT model. Now if $\nabla-\nabla_{\rm ad}$ from two models are vastly different, so is the
 entropy jump of both.  Hence, if a certain entropy of the interior convection zone and thus
 a certain radius of the convection zone shall be matched, a much
 more drastic change in $\alpha$ is required with the KMLT model to avoid a too large  entropy
 jump (see a detailled discussion in \citet{Mont04}).

\section{Convective velocity and entropy fluctuations}
\label{sec:Convective velocity and entropy fluctuations}
\subsection{Convective flux}
\label{sec:Convective flux}

Part of the mode excitation rates stems from the advection of turbulent entropy
fluctuations by turbulent motions  (the so-called ``entropy source term'').
This term scales -- see Paper~I -- as $\alpha_s^2  < s_t^2 >  u_0^2$ where $s_t$
represents the entropy fluctuations due to turbulent convection,  $\alpha_s \equiv
\left (\partial p / \partial s \right )_\rho$, $s$, $\rho$ and $p$ are respectively
the density, the entropy and the pressure, $<>$ denotes spatial and time average,
$u_0^2 \equiv 1/3 \, <{\vec u}^2>$, where $\vec u$ is the velocity field associated
with the turbulence. The factor $1/3$ arises from the simplifying assumption made in
Paper~I that the acoustic emission is injected into the p-mode isotropically in all
three directions.

The ``entropy source term'' scales as the square of the convective flux $F_c$. Indeed,
we show (see Appendix~A) that it scales as $ \left ( \ds \alpha_s/\rho_0 T_0 \right)^2
\, \left (\ds {\bar \Phi}/3 \right ) \, \ds F_c ^2$, where $T_0$ and $\rho_0$ are the mean
temperature and density, respectively, and $\bar \Phi$ is the spatially averaged anisotropy
factor which is defined as \citep{Gough77}:
\eqn{
\bar \Phi(r) \equiv  \frac{<u^2> }{<u_z ^2>}  \equiv \frac{v^2 }{w ^2} 
\label{eqn:phiz}
}
where $u_z$ is the vertical component of $u$ and $v$ and $w$ are defined as
$v^2 \equiv <u^2>$ and $w^2 \equiv <u_z^2>$, respectively.

For the CGM96 formulation, $F_c$ is computed according to Eqs.~(2)  and
(17)-(21) in \citet{Heiter02} and for the MLT treatment, it is calculated
according to Eq.~(14.118) in \citet[ Chap.~14]{Cox68}.

$F_c$ can be viewed as function of $\alpha$ and $\nabla$, $F_c=h(\nabla,\alpha)$,
where $h$ is given by the adopted model of convection.

{\it ECGM and EMLT models:}
for these two models only one mixing-length parameter is involved and $F_c$ can
directly be retrieved from $\alpha$ and $\nabla$.

{\it KCGM and KMLT models:}
in the outer region ($\tau<\tau_1$) as well as in the interior region ($\tau > \tau_2$),
values of $F_c$ can be directly retrieved from $\alpha$ and $\nabla$. In the
transition region ($\tau_1  <\tau < \tau_2$), however, we have to deal with two
functions for the convective flux: $F_c^{(\rm i)}=h(\nabla_{\rm i},\alpha_{\rm i})$,
the convective flux calculated as in the interior, and
$F_c^{(\rm a)}= h(\nabla_{\rm a},\alpha_{\rm a})$, the convective flux calculated
for the atmosphere. As a result of Eq.~(\ref{eqn:nabla:full_model}), the convective
flux $F_c$ of the model in the transition region can be related to $F_c^{(\rm i)}$
and $F_c^{(\rm a)}$ as follows:
\eqn{
F_c = \lambda(\tau) F_c^{(\rm a)} + (1-\lambda(\tau)) F_c^{(\rm i)}
\label{eqn:fc:full_model}
}
where $\lambda(\tau)$ is -- like $\beta(\tau)$ (Eq.~\ref{eqn:nabla:full_model}) --
a function of $\tau$ which ensures the continuity of the convective flux. This
function must decrease with $\tau$ and must fulfill $\lambda(\tau)=1$ for $\tau \leq \tau_1$
and $\lambda(\tau)=0$ for $\tau \geq \tau_2$. We point out that both $F_c^{(\rm i)}$
and $F_c^{(\rm a)}$ fulfill $F_c+F_{\rm rad} = L_\odot/ 4\pi r^2$ at any optical depth
$\tau$. Just as for $\beta(\tau)$, the choice for $\lambda(\tau)$ is rather arbitrary.
As for the case of $\nabla$ (Eq.~\ref{eqn:nabla:full_model}), we assume
$\lambda(\tau)=\beta(\tau) =\ds (\tau_2-\tau)/(\tau_2-\tau_1)$ for
$\tau_1 < \tau < \tau_2$.

Figure \ref{fig:fc:alpha:v} shows  $F_c$ as a function of $\tau$ for the KMLT and KCGM
models. 

\smallskip
Calculation of the driving by the entropy source term requires, in addition to the
convective flux ($F_c$), a model for the  mean anisotropy ($\bar \Phi$). In the
CGM96 model of convection the expressions for $v^2$ and $F_c$ do not depend explicitly
on the  mean anisotropy factor $\bar \Phi$. However, CGM96 adopt a model of anisotropy
which fixes the ratio $x \equiv k_h / k_v$, where  $k_h$ and $k_v$ are the horizontal
and vertical wavenumbers associated with the eddy  of wavenumber $k$ (note that
$k^2 = k_h^2 + k_v^2$). As a result of that model, for the largest scales $x=1/2$
while it increases quadratically with the total wavenumber $k$ for the smaller
scales.  Let us define $\Phi(k) \equiv u^2(k) / u_z^2(k)$, a {\sl $k$-dependent}
anisotropy factor.
For an incompressible fluid -- a property assumed by the models investigated here (see
also CM91) -- one can show that $ \Phi(k)= 1+ x^{-1}$.
Hence $x=1/2$  implies $ \Phi =3$, i.e. an isotropic velocity field. As a result of
its functional dependence on $k$, from the larger  scales ($k\sim k_0$) to the
smaller scales ($k \gg  k_0$)  $\Phi(k)$ decreases
in the CGM96 model  from $\sim 3$ towards  $\sim 1$.  However, we recall
that the model of stochastic excitation (MSE) we consider is basically isotropic.
The anisotropy is taken into account only at large scales through Gough's mean
anisotropy factor ($\bar \Phi$). Therefore, although CGM96's treatment adopts a model
where the anisotropy varies with $k$, we are left with the inconsistency that the
turbulent spectrum, $E(k)$, assumed for the MSE (see Sect.\ \ref{procedude}) is
isotropic along the turbulent cascade. A possible anisotropy is only be taken into
account at large scales through $\bar \Phi$. But as the modes are predominantly
excited by turbulent eddies with $k \sim k_0$, which carry most of the kinetic
energy, this approximation is not expected to affect significantly our prediction. 
Hence, we assume $\bar \Phi=3$ for the CGM96 models in the discussion given below.

\subsection{Convective velocity}
\label{sec:Convective velocity}

Driving  of the oscillation modes by the Reynolds term is proportional to 
$\rho_0 v ^4$.

For the CGM96's treatment, $v^2$ is computed according to Eq.~(88-90) of CGM96. 

The MLT approach provides a theoretical expression for $w^2$, the mean square of the
vertical component of the convective velocity. It can be related to the velocity $v$
by means of $\bar \Phi$ (Eq.~\ref{eqn:phiz}). BV's formulation of the MLT assumes $\bar \Phi=2$
\citep{Gough77}. Here, we compute $w^2$  according to Eq.~(14.110) of
\citet[ Chap.~14]{Cox68} and deduce $v$ from Eq.~(\ref{eqn:phiz}) with $\bar \Phi=2$.

Just as the convective flux, the convective velocity $v$ can be viewed as a function
of $\nabla$ and $\alpha$, $v=f(\nabla,\alpha)$.

{\it ECGM and EMLT models:}
as for the convective flux, $v$ can be retrieved from $\alpha$ and $\nabla$.

{\it KCGM and KMLT models:}
in the outer region (i.e. $\tau < \tau_1$) as well as in the inner region
($\tau > \tau_2$), only one $\alpha$ and one $\nabla$ are defined. In those regions
$v$ is computed as $v=f(\nabla,\alpha)$. However, in the transition region we must
deal with two different values of $\alpha$ (namely $\alpha_{\rm i}$ and $\nabla_{\rm i}$
 from the inner region and $\alpha_{\rm a}$ and $\nabla_{\rm a}$  from
the atmosphere) and $v$ is not a linear function of $\alpha$. We thus face the
difficulty to  define properly a convective velocity consistent with $F_{\rm c}$
(Eq.~\ref{eqn:fc:full_model}) in this region.

We proceed as follows: $\nabla(\tau)$ and $F_{\rm c}(\tau)$ are defined by
Eq.~(\ref{eqn:nabla:full_model}) and Eq.~(\ref{eqn:fc:full_model}), respectively.
Then, at fixed $\nabla$ and $\tau$, we define an \emph{equivalent mixing-length
parameter}, $\alpha^{*}$, such that $F_c= h(\nabla,\alpha^{*})$. This parameter is
variable with depth.  We next compute $v=f(\nabla(\tau),\alpha^{*})$ which hence is
consistent with $F_{\rm c}$ of Eq.~(\ref{eqn:fc:full_model}). 

\begin{figure}[ht]
        \begin{center}
          \includegraphics[width=8.5cm]{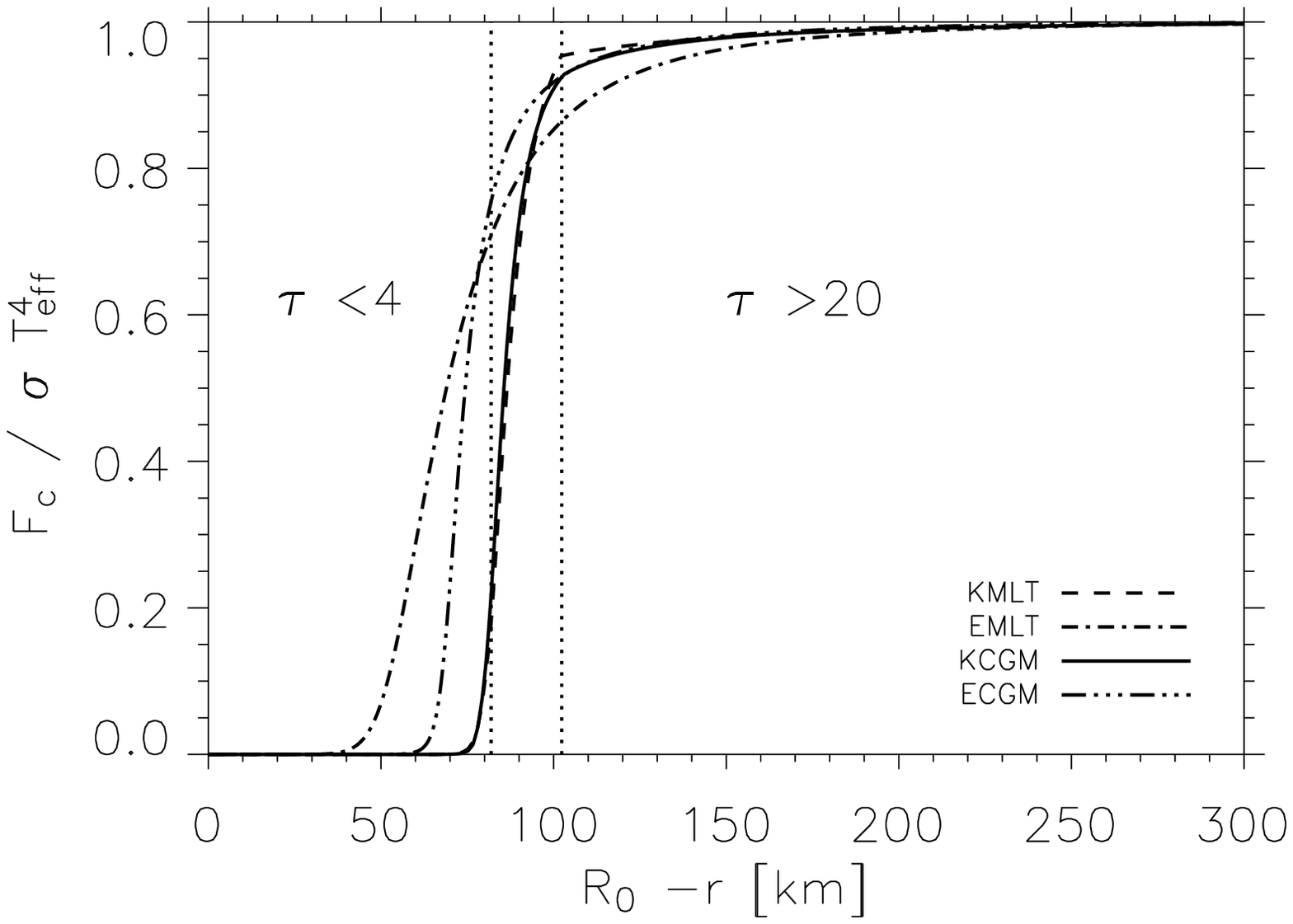}
	  \includegraphics[width=8.5cm]{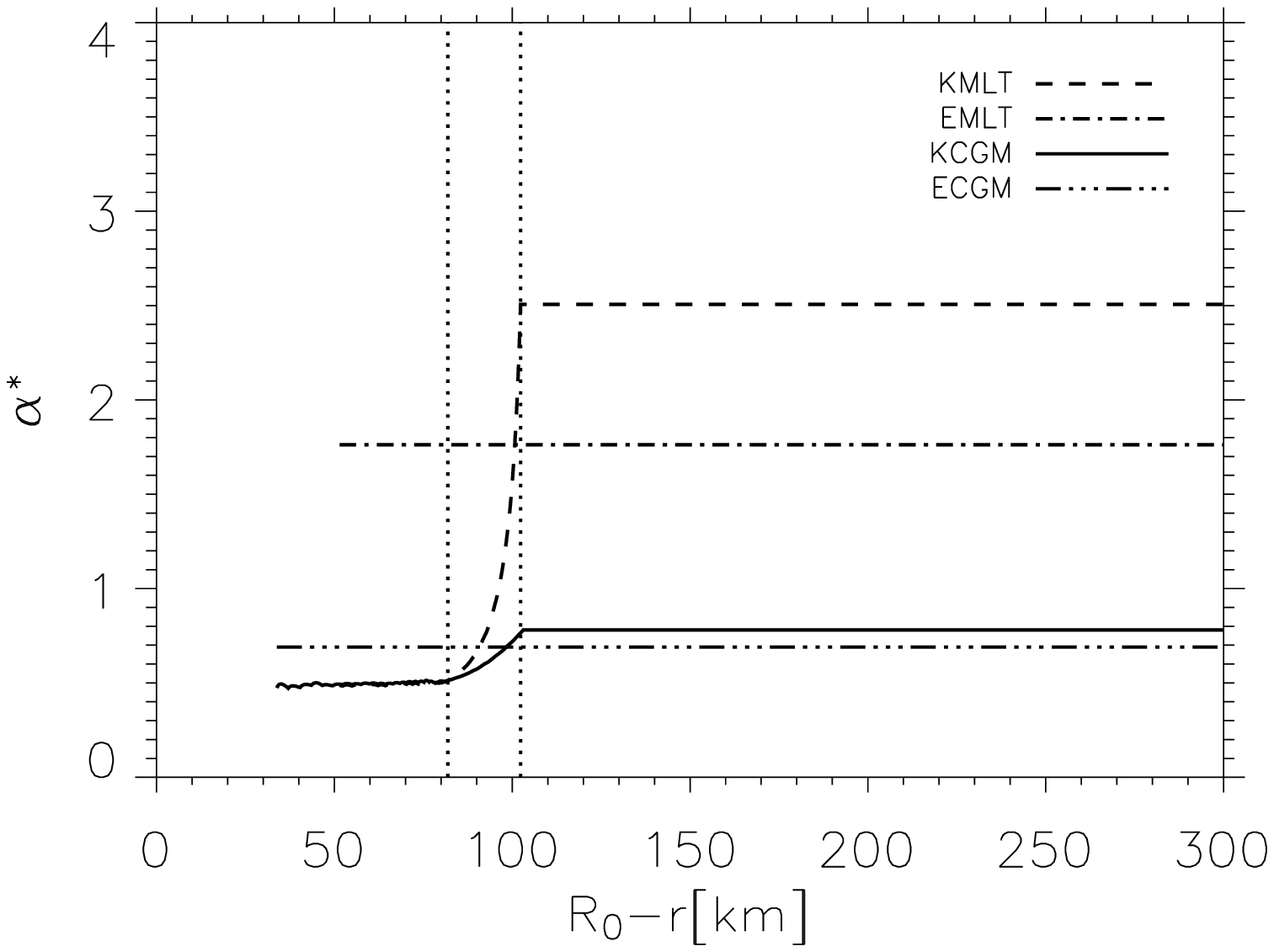}
	  \includegraphics[width=8.5cm]{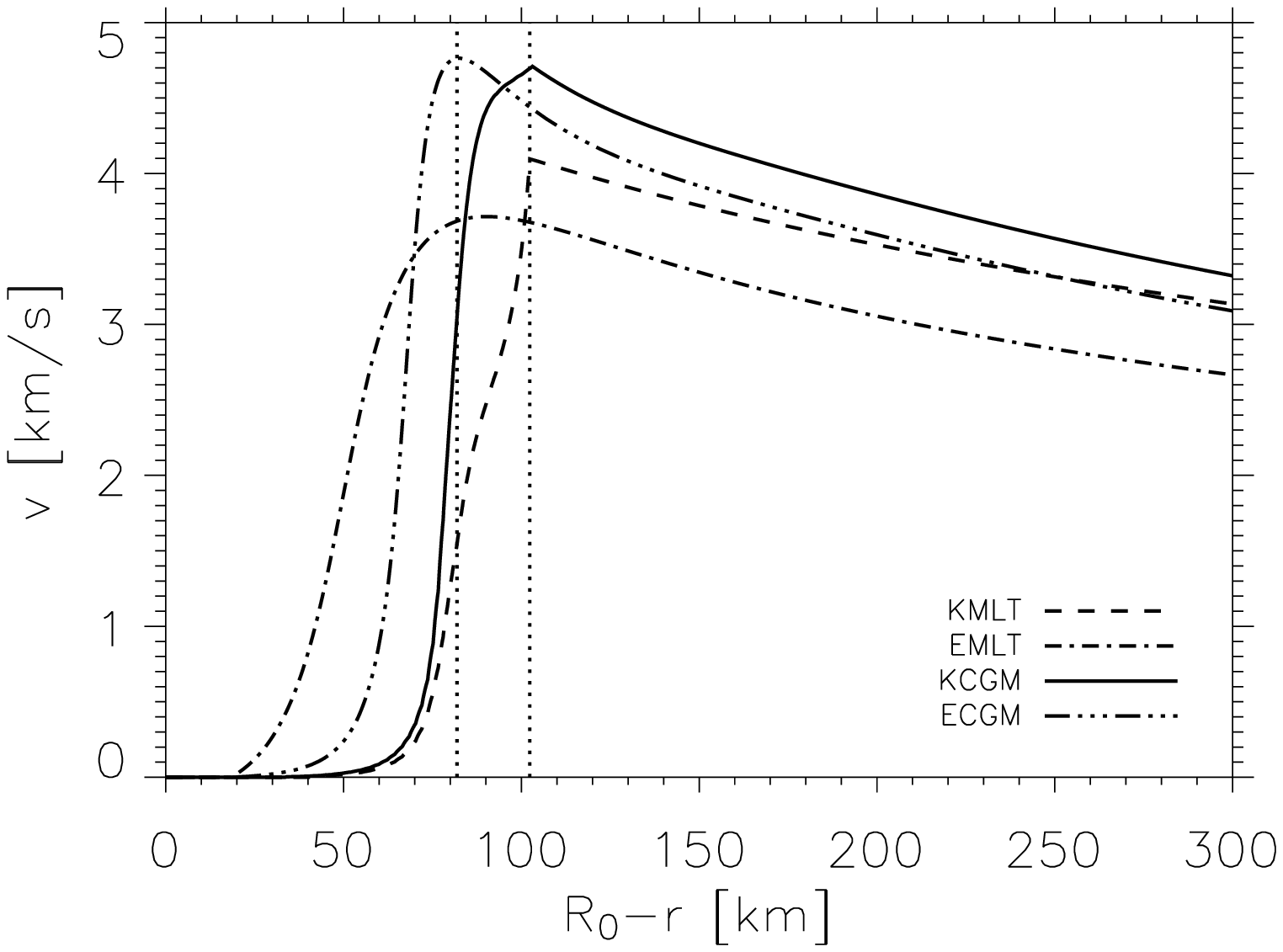}
        \end{center}    
        \caption{ {\bf Top:} $F_c$ is plotted versus $R_\odot -r$ for the KMLT model
		(dashed curve),  EMLT model (dot-dashed line), ECGM model (dot-dot-dot-dashed
		line) and KCGM model (solid line). As in Fig.~\ref{fig:grad:v:fc} the dotted
		vertical lines delimit the transition region.
{\bf Middle:} Same as the top panel for $\alpha^*$ (see Sect.~\ref{sec:Convective velocity}).
{\bf Bottom:} Same as the top panel for the root mean square of the convective velocity, $v$.
}
        \label{fig:fc:alpha:v}
        \end{figure}

Figure \ref{fig:fc:alpha:v} (middle and bottom) shows $\alpha^*$ and $v$ as a function
of depth. For the MLT models, our calculation of $v$ assumes $\bar \Phi=2$ which is
consistent with BV's formulation of the MLT \citep[see][]{Gough77} and for the CGM
models it assumes $\bar \Phi=3$. 


\subsection{Comments}
\label{sec:Convective flux and velocity:Comments}

As shown in Fig.~\ref{fig:fc:alpha:v}, the EMLT and KMLT models have very different
convective structures: up until close to the top of the superadiabatic region
(located at $R_\odot-r \sim 70$~km for those models), the EMLT model results in a
convective velocity and convective flux smaller than those of the KMLT model. This is
a consequence of the fact that $\alpha^{\rm EMLT} < \alpha_{\rm i}^{\rm KMLT}$ (see
Table~\ref{tab:kurucz} and Table~\ref{tab:edd}). On the other hand, the EMLT model
results in larger $v$ and $F_{\rm c}$ above the top of the superadiabatic region 
because $\alpha^{\rm EMLT} > \alpha_{\rm a}^{\rm KMLT}=0.5$.

In contrast with the MLT models, the KCGM and ECGM models have rather similar
convective structures. Indeed, $v^{\rm KCGM}$ and $v^{\rm ECGM}$
($F_{\rm c}^{\rm KCGM}$ and $F_{\rm c}^{\rm ECGM}$ resp.) have very similar shapes.
This in turn is a consequence of the fact that the CGM models require values of
$\alpha_i^{\rm KCGM}$ and of $\alpha^{\rm ECGM}$ close to that one required for the
atmosphere ($\alpha_a=0.5$).

For the KMLT model, there is a pronounced ``kink'' at the bottom boundary of the
transition region ({\it i.e.} at $\tau =20$ or $R_\odot-r = 100$~km),
especially for $v$. For the KCGM model the  ``kink'' is much less important. These
features are directly connected with the angular point observed for $\nabla$ in that
layer (see Sect.~\ref{sect:The solar models:Comments} and Fig.~\ref{fig:grad:v:fc}).

At the bottom of the transition region ({\it i.e.} at $\tau=\tau_2$), for both
the KMLT and KCGM models, $\alpha^*(\tau)$ reaches -- as expected --  the ``interior''
value of the mixing-length parameter ({\it i.e.} $\alpha_{\rm i}$), namely: 
$\alpha_{\rm i}^{\rm KCGM} = 0.78$ and $\alpha_{\rm i}^{\rm KMLT} = 2.51$.
At the top of  the transition region ({\it i.e.} at $\tau=\tau_1$), $\alpha^*(\tau)$
reaches for both the KCGM and KMLT models the asymptotic value $\alpha^* \simeq 0.50$.

\section{Calculation of the solar $p$~mode excitation rates}

\subsection{Procedure}
\label{procedude}

We compute the rate $P$ at which acoustic energy is injected into solar radial
$p$-modes according to the theoretical model of stochastic excitation of Paper~I and
assume here -- as in Paper~II and Paper~III -- that injection of acoustic energy into
the modes is isotropic and consider only radial p~modes. The rate at which a given
mode with frequency $\omega_0=2\pi \nu_0$ is excited is then calculated with the set
of Eqs~(1-11) of Paper~III and Eq.~(3) of \citet{Samadi04}. 

The calculation requires the knowledge of three different types of quantities. First
of all, quantities which are related to the average properties of the medium:
\begin{itemize}
\item[$\bullet$] the mean density $\rho$, $\alpha_s$ (Eq.~\ref{eqn:alpha_s}), the mean
square convective velocity $v^2$, and the mean square of the entropy fluctuations $s^2$
(Eq.~\ref{eqn:s2}). They are obtained from the equilibrium models as explained in
Sect.~\ref{sec:Convective velocity and entropy fluctuations}.
\end{itemize}
Secondly, quantities which are related to the oscillation modes:
\begin{itemize}
\item[$\bullet$] the eigenfunctions $\xi_r$ and the eigenfrequency $\nu$. They are
computed with the adiabatic pulsation code FILOU \citep{Tran95} for each model. 
\end{itemize}
Finally, quantities which are related to the properties of the turbulent flow:
\begin{itemize}
\item[$\bullet$] the wavenumber ($k$) dependency of $E$, i.e.\
the turbulent kinetic energy spectrum;
\item[$\bullet$] the values and depth dependency of $k_0$, 
the wavenumber at which the convective energy has its maximum and is ``injected''
into the inertial range of the turbulent kinetic energy spectrum $E$;
\item[$\bullet$] the $\nu$-dependency of $\chi_k$, the frequency component 
of the  auto-correlation product  of the turbulent velocity field.
\end{itemize}

According to the results in Paper~II and Paper~III obtained on the basis of a 3D
numerical solar granulation simulation, the $k$-dependency of $E(k,\nu)$ is
approximately reproduced by an analytical spectrum called `Extended Kolmogorov
Spectrum' (EKS) and defined in \citet{Musielak94}. The $\nu$-dependency of $\chi_k$
is found to be better modelled with a Lorentzian function rather than by a Gaussian
function, which is usually assumed for $\chi_k$ (see Paper~III). Within most part of
the excitation region the spatially averaged anisotropy factor $\bar \Phi$ is found almost
constant with a value of $\bar \Phi \sim 2$ in agreement with BV's value.

At the top of the superadiabatic region, it was found that $k_0 \sim
3.6 ~{\rm Mm}^{-1}$ and that $k_0$ decreases slowly inwards with depth (see Paper~II).
A good approximation for our region of interest within the sun is to assume
a constant $k_0$.

\subsection{Comparison with the observations}

\begin{figure}[ht]
       \begin{center}
        \includegraphics[width=\lenA]{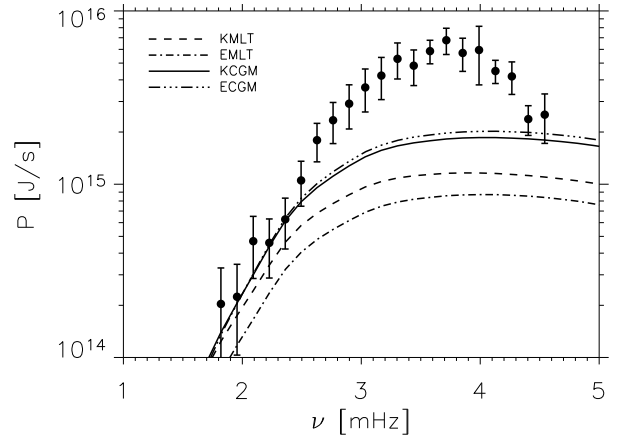}
       \end{center}    
        \caption{The computed  solar $p$~mode excitation rates, $P(\nu)$, are plotted
		versus the frequency for the KMLT (dashed line), EMLT (dot-dashed line), ECGM
		(dot-dot-dot-dashed line) and KCGM (solid line) models. The filled circles
		represent the `observed'  solar values of $P(\nu)$ derived -- according to
		Eq.~(\ref{eqn:P_vs2}) -- from the amplitudes and line widths of the $\ell=0$ 
        $p$-modes measured by \citet{Chaplin98}.}
\label{fig:pow}
        \end{figure}

Results for $P$ are presented in Fig.~\ref{fig:pow}. The theoretical estimates for
$P$ are compared with the `observed' $P_{\rm obs}$, calculated from
\citet{Chaplin98}'s seismic data according to the relation:
\eqn{
     P_{\rm obs}  (\nu) = 2 \pi \, \Gamma \, \frac{ I  }{\xi_{\rm r}^2(r_s) } \, 
     v_s^2 (\nu) =  2 \pi \, \Gamma \,  {\mathcal M }  \,  v_s^2 (\nu),
\label{eqn:P_vs2}
}
where $\nu$ is the mode frequency, $r_s$ is the radius at which oscillations are
measured,   
\eqn{
I \equiv   \int_0^{M} dm \,   \xi_{\rm r}^2 
\label{eqn:I} 
} 
is the mode inertia, ${\mathcal M }= I / \xi_{\rm r}^2(r_s)$ is the mode mass, and
finally $\Gamma$ and $v_s$ are the mode line-width and the mode surface velocity,
respectively, and are obtained from \citet{Chaplin98}. We point out that, according to
the definition of Eq.~(\ref{eqn:P_vs2}), the derived value of $P_{\rm obs}$ depends on
the model one considers through ${\mathcal M }$. Indeed, for a given mode,
${\mathcal M }$ -- {\it a~priori} -- differs from one model to another. However, the
mode masses of the models we consider here are very close to each other such that the
changes on $P_{\rm obs}$ due to the use of the mode mass of the different models are
not significant compared to the error bars attached to the measurements. For each mode,
we can then compute a unique value for $P_{\rm obs}$ and compare the values of $P$
obtained for the different models to each other and to  $P_{\rm obs}$. We choose to
derive ${\mathcal M} $ from the radial eigenfunctions $\xi_r$ computed for the KCGM
model and we adopt  $r_s = R_\odot + 200$~km  consistently with \citet{Chaplin98}'s
observations.    
 
As shown in Fig.~\ref{fig:pow}, differences between $P_{\rm KCGM}$ and $P_{\rm ECGM}$
are found smaller than the error bars associated with $P_{\rm obs}$ from
\citet{Chaplin98}. This is a consequence of the fact that the ECGM and KCGM models
present very similar convective  structures (see Sect.~\ref{sec:Convective flux and
velocity:Comments} and Fig.~\ref{fig:fc:alpha:v}). On the other hand, $P_{\rm KMLT}$
is found significantly larger than $P_{\rm EMLT}$ as a consequence of the fact that
the KMLT models result in larger $v$ and $F_{\rm c}$ values than the EMLT ones (see
Sect.~\ref{sec:Convective flux and velocity:Comments} and Fig.~\ref{fig:fc:alpha:v})
for $\tau \gtrsim 10\dots 20$.

Furthermore, $P_{\rm KCGM}$ and $P_{\rm ECGM}$ are found closer to $P_{\rm obs}$ than
$P_{\rm KMLT}$. However, above $\nu \gtrsim 2.5$~mHz, differences between $P_{\rm obs}$
and $P_{\rm KCGM}$ (or $P_{\rm ECGM}$) remain important. The origin of this discrepancy
is discussed in Sect.~\ref{conclusion}.

For the KCGM and ECGM models we have so far assumed $\bar \Phi=3$. According
to Eq.~(\ref{eqn:s2}), assuming $\bar \Phi=2$ -- a value which is consistent with
results from the 3D solar simulation -- results in a driving by the entropy
source smaller by a factor $\sim$ 2/3 compared to the case $\bar \Phi=3$. This
decrease, however, remains  small compared to the difference in $P$ between
the different models and hence does not influence our main results.

     \begin{figure}[ht]
       \begin{center}
        \includegraphics[width=\lenA]  {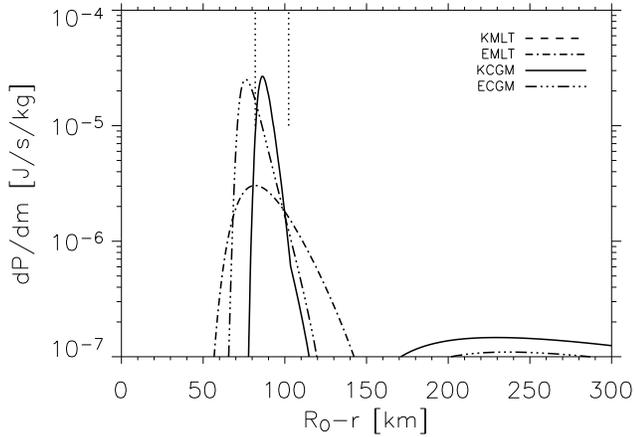}
       \end{center}    
        \caption{ The integrand $dP/dm$ (Eq.~(2) of Paper~II) is plotted versus 
		$R_\odot - r$ for a mode with order $n=20$ ($n=1$ being the fundamental mode)
	    for the KMLT model (dashed line), EMLT model (dot-dashed line), ECGM model
		(dot-dot-dot-dashed line) and KCGM model (solid line).}
\label{fig:dP_dm}
        \end{figure} 

We present in Fig.~\ref{fig:dP_dm} the integrand $dP/dm$ (Eq.~(2) of Paper~II)
corresponding to the excitation rate $P$ of a mode of order $n=20$. The plot
is done for the four solar models. For three of the four models the extent of the
region where most of the excitation takes place is very thin (less than 50~km).
 Obviously, this is the consequence of the very shallow extent
of the superadiabatic region (see Fig.~\ref{fig:grad:v:fc}). This tiny extent
of the excitation region strongly contrasts with that one found in
\citet{Stein01II} and in Paper~III. For instance, the latter authors found
a size of order $\sim 500$~km for a mode of the same order. This discrepancy
is attributed to the local nature of the convective treatments
investigated in this work (see Sect.~\ref{conclusion}). Note that in
Fig.~\ref{fig:dP_dm} the EMLT model with its large value of $\alpha$ for the entire
superadiabatic region predicts a broader excitation region than the other
models. This is due to a smaller superadiabatic peak. However, in this model the
transport of convective heat occurs with a smaller convective velocity and hence,
as expected, the excitation is smaller than for the other models. We note here
that despite the EMLT model results in a temperature structure closer to the
numerical simulations (smaller superadiabatic peak), the predicted excitation
amplitudes are smaller than for the other models investigated here and are the
most discrepant in comparison with the data. This confirms that excitation
rates provide a decisive additional test for convection models.

\section{Conclusions}
\label{conclusion}

We have computed the rates $P$ at which acoustic energy is injected into the solar
radial $p$~modes for several solar models. The solar models are computed with two
different local treatments of convection: the classical mixing-length theory (MLT
hereafter) and \citet[][ CGM96 hereafter]{Canuto96}'s formulation.

For one set of solar models (EMLT and ECGM models), the atmosphere is gray and assumes
Eddington's approximation. The models only assume one mixing-length parameter value and reproduce
the solar radius at solar age but not the Balmer line profiles. For a second set of models
(KMLT and KCGM models), the atmosphere is built using a $T(\tau)$ law which has been
obtained from a Kurucz's model atmosphere computed with the same local treatment of
convection. The mixing-length parameter in the model atmosphere is chosen so as to provide
a good agreement between synthetic and observed Balmer line profiles, while the
mixing-length parameter in the interior model is calibrated so that the model reproduces
the solar radius at solar age.

Both the KMLT and the KCGM models reproduce the Balmer line profile and the solar
radius and luminosity but -- as shown in Figs. (2)-(4) -- the CGM models model the
transition between the region of high convective efficiency (the interior) and the
region of low efficiency (the atmosphere) in a more realistic way than the
MLT models, as they reproduce the observed excitation rate $P$ more closely and
predict a smoother transition region.
Furthermore the KMLT model requires a change of the mixing-length  by a factor of five in a layer of $\sim 20$ km thickness, which is significantly less than a
pressure scale height ($\sim$~300 km). Given the meaning of alpha, this means
that the mixing-length varies from about 150 km to 750 km in a layer
of $\sim 20$~km {\em thickness (!)}, which does not make much sense from a physical
 point of view.  On the other hand, the KCGM model does not require such large change
in $\alpha$.

For the MLT treatment, the oscillation excitation rates, $P$, do significantly depend
on the properties of the atmospheres investigated here. Indeed, differences in $P$
between the EMLT model and the KMLT model are found to be very large. On the other
hand, for the CGM96 treatment, differences in $P$ between the ECGM and the KCGM models
are very small compared to the error bars attached to the seismic measurements. 
This result shows that an Eddington gray atmosphere can be assumed for the
calculation of $P$ when the CGM96 formulation is adopted. This would be particulary
convenient in the case of massive computations of $P$ for a large set of stellar models.

For the EMLT and KMLT models, $P$ is significantly underestimated compared to the solar
seismic constraints obtained from \citet{Chaplin98}'s measurements. KCGM and ECGM
models yield values for $P$ closer to the seismic data than the EMLT and KMLT models.  
Contraints on  the ${\rm H}_\beta$ Balmer line profile and on the solar radius can be
satisfied by the two formulations (MLT and CGM96), provided that the mixing-length 
parameters $\alpha_{\rm i}$ and $\alpha_{\rm a}$ are suitably adjusted. Once the above
constraints are satisfied, the solar seismic data provide additional valuable
constraints and according to the present investigation (focussed on local approaches)
they cleary favor the CGM96 treatment.

\smallskip

Our calculations are based here on \citet{GN93}'s solar abundances.
There was recently a change in these values (see \citet{Asplund04,Asplund05}) with quite some implications for the internal structure of the Sun (see \citet{Basu04,Montalban04b,Bahcall05,Antia05}). 
Then the implication for the calculation of the excitation rates ($P$) must be in principe tested consistently by changing the solar mixture both in the interior and the atmosphere (work in progress). However, we  except small changes in $P$. 
Indeed, as a first step we have calculated a solar model having the low metallicity $Z/X=0.0171$ inferred from the new \citet{Asplund04}'s revised solar abundances in which the interior calculation is based on the detailed Asplund et al's mixture and associated opacities while the atmosphere considers the low metallicity but keeps the \citet{GN93}'s solar abundances and associated opacities.
Changes in $P$ smaller than $\sim$ 10\% ---~  much smaller than the errors bars ($\sim$ 20\%) associated with the current measurements ~--- have been obtained.

\smallskip

The remaining discrepancy above $\nu \gtrsim 2.5$~mHz between computed and observed $P$
(Fig.~\ref{fig:pow}) is attributed to the assumption of locality in the present treatment
of convection. As a matter of fact, \citet[][ Paper~III]{Samadi02II} have succeeded
in reproducing the seismic constraints much better using constraints from a 3D
solar granulation simulation. One reason for this improvement is that convection is
intrinsically a non-local phenomenon. In the terminology of classical turbulence
modelling, the eddies located at different layers contribute to the convective flux of
a given layer \citep[cf.\ also the discussion in][]{Houdek96}. Hence, a non-local
description of convection is expected to predict a more extended superadiabatic region.
This is suggested, for instance, by the comparison of our present results with
that of Paper~II. Non-local theories --~such as those by \citet{Gough77} and by
\citet{Canuto98}~-- also support this explanation: they typically predict
a smaller temperature gradient in the superadiabatic region than the local
theories do \citep[see][]{Houdek96,Kupka02a} and thus a more extended
superadiabatic region as well.  Another property of solar granulation
caused by non-locality is the observed asymmetry between the areas covered by
up- and downflows. This allows for a lower root mean square velocity while
larger velocities (and thus more effective mode driving) can be reached in
the downflows with their much higher velocities (note that such an asymmetry
can be accounted for through non-local models as proposed by \citet[][ see
also \citealt{Kupka02a}]{Canuto98}, although in a more simplified manner). 
 On the other hand, the local models
studied here cannot account for the different properties of up- and downflow
areas at all. Hence, the superadiabatic region in these models is physically
different from that one expected from non-local models and found 
in numerical simulations. Solar modes above $\nu \gtrsim 2$~mHz, however,
are predominantly excited in the superadiabatic region (Paper~III). A larger
amount of acoustic energy is then injected in those modes when convection
is treated -- as is in the 3D simulation or on the base of a non-local
theory -- in a more realistic manner than through local theories. 

\smallskip

The results presented here so far only concern our Sun. \citet{Samadi02b} found that
$P$ scales as $(L/M)^s$ where $s$ is the slope of the power law and $L$ and $M$ are
the mass and luminosity of their computed 1D stellar models. By building a set of
stellar models with the MLT and another one with \citet{Gough77}'s non-local
formulation of convection, the authors found that the slope $s$ is quite sensitive to
the 1D treatment of convection. In this respect, it would be interesting to compare the
influence of the CGM96 formulation or of \citet{Canuto98}'s non-local convection
treatment on the value of $s$ and to test whether or not measurements of $P$ can
-- for stars other than the Sun -- discriminate the best 1D treatment of convection.


\begin{acknowledgements}
RS's work has been supported in part by by the Particle Physics and Astronomy
Research Council of the UK under grant PPA/G/O/1998/00576  (who also supported
FK), by the Soci\'et\'e de Secours des Amis des Sciences (Paris, France)
and by  Fundac\~ao para a Ci\^encia e a Tecnologia (Portugal) under grant
SFRH/BPD/11663/2002. We thank  Achim Weiss for having carefully read this manuscript and R. Kurucz for having given access to the ATLAS9 code and associated data. Finally we thank the referee for his judicious suggestions. 
\end{acknowledgements}

\appendix
\section{Relation between the entropy fluctuations and the convective flux}
As in Paper~I, we relate the entropy fluctuations, $s_t$, to temperature fluctuations,
$\theta$, as follows:
\eqn{
 s_t = \frac{c_p}{T_0}  \, \theta
\label{eqn:s_t:theta}
}
where $T_0$ is the mean temperature and  $c_p =  (\partial s / \partial \ln T)_p$. 
Hence, the mean square of the entropy fluctuations, $< s_t^2>$, can be expressed as 
\eqn{
 < s_t^2> \approx \left(\frac{c_p}{T_0} \right)^{\! 2} \, <\theta^2>.
\label{eqn:s_t:theta:ms}
}
The convective flux is related to $\theta$ and $u_z$, the vertical component of
convective velocity, as 
\eqn
{
F_c \approx \rho_0 \, c_p \, <u_z \, \theta>,
\label{eqn:Fc}
}
where $\rho_0$ is the mean density.
We furthermore assume, consistently with the adopted quasi-normal approximation in
Paper~I, that $ <u_z \, s_t>^2$ can be decomposed as
\eqn{
 <u_z \, s_t>^2 \, = \, <u_z^2> \, <s_t^2> \, =\, w^2 \, <s_t^2>.
\label{eqn:ws2}
}
Finally, one can show that
\eqn{
\alpha_s \equiv \left (\derivp{p}{s} \right )_\rho = \rho_0 T_0 \Gamma_1
\nabla_{\rm ad},
\label{eqn:alpha_s}
}
where $s$ is the entropy and $p$ the pressure, $T_0$ is the mean temperature,
$\displaystyle{\Gamma_1 = \left( \partial \ln p / \partial \ln \rho \right )_s }$
is the adiabatic exponent and $\nabla_{\rm ad} = (\partial \ln T/ \partial \ln p)_s$
is the adiabatic temperature gradient. 

The mean square of the entropy fluctuations can  then be deduced from the set of
Eqs.~(\ref{eqn:s_t:theta}-\ref{eqn:alpha_s}) and Eq.~(\ref{eqn:phiz}):
\eqn{
<s_t^2> \approx \frac{ \bar \Phi}{3} \, \left( \frac{F_c}{ \rho_0 T_0 u_0  } \right)^2,
\label{eqn:s2}
}
where $\bar \Phi$ is the spatially averaged anisotropy factor, which is defined in
Eq.(\ref{eqn:phiz}). Driving by the entropy source term is then proportional to
$\left( \ds \alpha_s/\rho_0 T_0 \right)^2 \, \left (\ds  \bar \Phi/3 \right ) \, \ds F_c ^2$
-- see Paper~I -- and thus scales like the square of $F_c$.


\bibliographystyle{aa}

\end{document}